\documentclass[twocolumn,aps,prl,superscriptaddress]{revtex4-1}

\usepackage{dcolumn}
\usepackage{amsmath,amssymb}

\usepackage{bm}
\usepackage{bbm}
\usepackage{color}
\usepackage{overpic}
\usepackage{latexsym}
\usepackage{epstopdf}
\usepackage{color}
\usepackage[english]{babel}
\usepackage{times}
\usepackage{latexsym}
\usepackage{psfrag,graphicx}
\usepackage{epsf}
\usepackage{subfigure}
\usepackage{amsmath}
\usepackage{amssymb}
\usepackage{amsfonts}
\usepackage{epstopdf}
\usepackage{appendix}
\usepackage{enumitem}
\usepackage{braket} 				
\usepackage{amsthm}					
\usepackage{soul}

\definecolor{mygrey}{gray}{0.35}
\definecolor{myblue}{rgb}{0.2,0.2,0.8}
\definecolor{myzard}{cmyk}{0,0,0.05,0}
\definecolor{mywhite}{rgb}{1,1,1}
\definecolor{mygreen}{rgb}{0,.6,0}
\definecolor{myred}{rgb}{0.9,0.1,0.}
\usepackage[colorlinks=true,citecolor=myblue,linkcolor=myblue,urlcolor=myblue]{hyperref}

\newcommand{\BTF}[1]{{{#1}}}

\begin{document}

\title{
{Limited-control metrology approaching the Heisenberg limit without entanglement preparation}}

\author{Benedikt Tratzmiller}
\affiliation{Institut f\"ur Theoretische Physik and IQST, Albert-Einstein-Allee 11, Universit\"at Ulm, D-89081 Ulm, Germany}

\author {Qiong Chen}
\affiliation{Institut f\"ur Theoretische Physik and IQST, Albert-Einstein-Allee 11, Universit\"at Ulm, D-89081 Ulm, Germany}
\affiliation{College of Physics and Electronics, Hunan Normal University, Changsha 410081, China}

\author {Ilai Schwartz}
\affiliation{Institut f\"ur Theoretische Physik and IQST, Albert-Einstein-Allee 11, Universit\"at Ulm, D-89081 Ulm, Germany}
\affiliation{NVision Imaging Technologies GmbH, Albert-Einstein-Allee 11, Universit\"at Ulm, D-89081 Ulm, Germany}

\author {Susana F. Huelga}
\affiliation{Institut f\"ur Theoretische Physik and IQST, Albert-Einstein-Allee 11, Universit\"at Ulm, D-89081 Ulm, Germany}

\author {Martin B. Plenio}
\affiliation{Institut f\"ur Theoretische Physik and IQST, Albert-Einstein-Allee 11, Universit\"at Ulm, D-89081 Ulm, Germany}

\date{\today}

\begin{abstract}
{Current metrological bounds typically assume full control over all particles that are involved in the protocol.
Relaxing this assumption we study metrological performance when} only limited control is available. As an example,
{we measure} a static magnetic field when a fully controlled quantum sensor is supplemented by particles
over which only global control is possible. We show that even {for a noisy quantum sensor, a protocol that maps the magnetic field to a precession frequency can achieve transient super-Heisenberg scaling} {in measurement time 
and Heisenberg scaling in the particle number. This leads to} an {estimation uncertainty} that approaches 
that {achievable under} full control to within a factor independent of the particle number {for a given 
total time}. {Applications to hybrid sensing devices and the crucial role of the quantum character of the 
sensor are discussed.}
\end{abstract}

\maketitle
\section{Introduction} The use of quantum resources in sensing and metrology has a longstanding
history which originated with the use of single-mode squeezed states \cite{Caves81}
and multi-particle spin-squeezing \cite{WinelandBI+92,WinelandBI+94}, i.e., entanglement,
{to enhance precision in interferometry and atomic spectroscopy.}

The {goal of quantum metrology} is the optimisation of the scaling of metrological
precision with the available physical resources {\cite{Giovannetti, Braun}}. Notably,
in a noiseless setting, independent preparation and measurement of {$M$ particles in
parallel results} in a {$1/\sqrt{M}$} scaling of the uncertainty{, the so called
standard quantum limit (SQL)}, while the collective preparation of the particles in
an entangled state leads to a {$1/M$}-scaling, commonly referred to as Heisenberg
scaling (HS) \cite{WinelandBI+92,WinelandBI+94} {(see \cite{hyllus2012fisher, toth2013extremal}
for more general upper bounds obtained via the quantum Fisher information). The use of
entangled states is necessary to achieve the optimal precision and exact
HS but sequences of probe states with an asymptotically vanishing amount
of entanglement can reach a scaling arbitrarily close to the Heisenberg limit (HL)
\cite{augusiak2016asymptotic}.} Environmental noise is known to have a non-trivial
impact on metrology \cite{HuelgaMP+97} and a meaningful comparison of different
schemes needs to specify carefully the conditions under which the metrological protocol
is carried out, such as the number of particles or the total amount of time available
\cite{HuelgaMP+97, EscherMD11}. A wide variety of setting has been analysed \cite{HuelgaMP+97,EscherMD11,DemkowiczKG12,haase2018fundamental} and noise models have
been found to result in metrological scaling intermediate between SQL and the
HL \cite{ChinHP12, matsuzaki2011magnetic, macieszczak2015zeno, smirne2016ultimate, haase2018fundamental}.
However, these results depend on access to perfect and {arbitrarily} fast control and feedback
operations \cite{SekatskiSK+17, DD2014}.

In practice, however, only limited control is possible over experimental resources and the
asymptotic regime {of large} numbers of fully controlled particles is not accessible. 
{What can be achieved in metrology for systems where, for example, particles cannot be addressed individually, multi-particle
	quantum gates are not available or the rate of measurements, feedback and the number of accessible particles
	is limited?}

In order to initiate investigations of this type in a concrete setting, we allow ourselves to
be motivated by the recently developed concept of quantum-hybrid sensors \cite{CaiJP14,WrachtrupF14}.
These are devices that integrate at least two components, one being {a fully controlled} quantum
sensor and another, typically an assembly of quantum particles, mutually interacting or
not, that are coupled to the quantum sensor but over which there is no individual control.
This second component acts as a transducer of a signal to a form that is then detected
by the quantum sensor.
{An example is a device composed of a piezo-magnetic material deposited on a diamond surface
that translates a force into a stray magnetic field which is then detected by a shallowly implanted
nitrogen-vacancy (NV) center in diamond\cite{GruberDT+97,MullerKC+14,WuJP+16}.
Another possibility, motivated by recently realised nanoscale NMR measurements
\cite{SchmittGS+2017,BossCZ+17,GlennBL+18}, consists of an ensemble of $M$ nuclear
spins and an NV center. Here an applied magnetic field can be measured as it induces
a nuclear Larmor precession which can be monitored by an NV center. Similarly, atoms 
in microfabricated vapour cells \cite{ShahKS+07} could allow for observation of their Larmor 
precession by an NV center rather than a classical laser field.
We assume that the quantum sensor is subject to noise,
and cannot exert individual control over the noise-free auxiliary spins.}

While we are motivated by concrete settings, our analysis yields more compact expressions,
without affecting the scaling properties, by assuming that all involved spins have the same magnetic
moment. Furthermore, all the following considerations will neglect direct interactions between the
auxiliary spins.

By means of analyzing the scaling of the Fisher information with respect to particle number $M$ and measurement time $T$ we will show that, despite the limitations of partial control, it is possible
to get close to the Heisenberg limit, we will discuss the origin of this scaling and compare our scheme to a single quantum sensor without auxiliary spins.

\begin{figure}
	\centering
	\includegraphics[width=.5\textwidth]{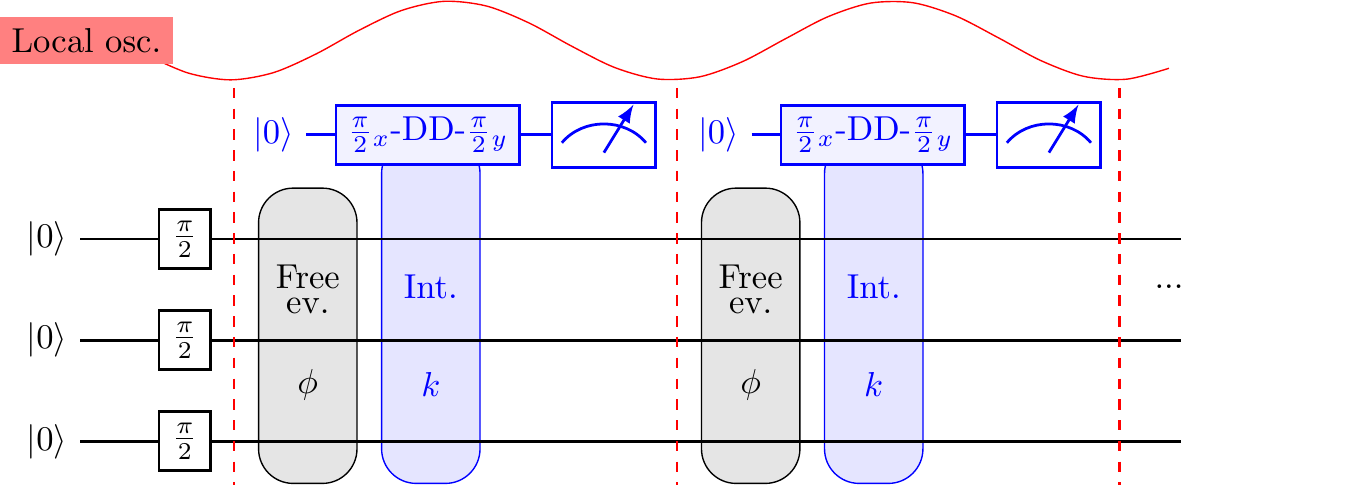}
	\caption{
The proposed measurement scheme uses a control sequence on the sensor spin (blue) to weakly measure the auxiliary spins (black) without the need of further control after initialisation in either the pure state $\ket{+++...}$ as shown here or the completely mixed state. {The weak measurement is realised by initialising a sensor spin (blue) into $\ket{0}$, applying $\pi/2$-pulses and a dynamical decoupling (DD) sequence to weakly entangle the sensor spin with the auxiliary spins and finally measuring the sensor spin, applying an effective CPTP map onto the auxiliary spins.} In between these measurements {that are synchronised with an external local oscillator,} the auxiliary spins acquire a phase $\phi \propto {\mu_{n}} B$, leading to a total phase $n \phi$ after $n$ cycles. }
	\label{Fig0}
\end{figure}

\section{Approaching the Heisenberg limit without entanglement preparation}

{Before stating the main results, we briefly recapitulate the achievable uncertainties under unconstrained metrology to make it available for later comparison with our schemes.}

\subsection{The ideal case of full quantum control} {Consider} a quantum sensor and $M$ auxiliary
spins, all with the same magnetic moment {${\mu_{n}}$, over which we can exert arbitrary and fast control.
Then the optimal uncertainty for the estimation of the magnetic field in a time $T$ in the absence of
noise is obtained via Ramsey spectroscopy using the $M+1$ accessible particles prepared in a {highly
entangled} state of the form $(|0\ldots 0\rangle + |1\ldots 1\rangle)/\sqrt{2}$ and is given by} \cite{HuelgaMP+97}
\begin{equation}
    \Delta B = \frac{\hbar}{{\mu_{n}} (M+1) T}.
    \label{Heisenberg}
\end{equation}
We observe a linear decrease in the uncertainty, i.e. HS, both in the total measurement time $T$ and the number of spins $M$.

\subsection{ {The case} of limited control setting} We consider a perfectly controlled quantum
sensor supplemented by $M$ auxiliary spins all having the same magnetic moment. These $M$ auxiliary
spins can be controlled by a global field and interact weakly with the quantum sensor, i.e. the product
of sensor-auxiliary spin interaction strength and interaction time is much smaller than unity.
For simplicity we assume that each auxiliary particle is interacting with the same strength and phase
with the quantum sensor, however the basic findings remain the same in the more general case. 

{We measure a static external magnetic field $B$ with this hybrid sensor and determine
the achievable uncertainty $\Delta B$.} To this end we initialise the {auxiliary spins}
in a fully polarised state \cite{FN1} and then subject {them} to a $\pi/2$-pulse.
We determine the resulting rate of precession of these $M$ spins by periodically
measuring the time-dependent magnetic field generated by the precessing {spins}
in regular time intervals using the quantum sensor {and a dynamical decoupling {(DD)} sequence to weakly
entangle it with the auxiliary spins}. 
This allows for comparison of the precession frequency
	to that of a local oscillator \cite{SchmittGS+2017,BossCZ+17} as sketched in figure \ref{Fig0}.
{This DD sequence does not only cancel static noise on the sensor spin, but also allows to accumulate signal over several auxiliary spin Larmor periods by creating an effective $\sigma_z \otimes \sigma_x$ interaction. In order to see this, we
 start with a Hamiltonian of the form
	
	\begin{align}
	H = \frac{\Omega(t)}{2} \sigma_x &+ \sum_{m=1}^{M} \frac{\omega_L}{2} \sigma_z^{(m)} 
	\\& + A_\perp^{(m)} \sigma_{z} \otimes  \left(\sigma_x^{(m)} \cos({\phi_{0m}}) 
	 +  \sigma_y^{(m)} \sin({\phi_{0m}})  \right)\nonumber
	\end{align}
	that describes a NV center coupled to $M$ nuclear spins in the interaction picture
	where $\Omega(t)$ is the Rabi frequency with $\Omega(t)=0$ during the free evolution, $\omega_L$ is the nuclear Larmor frequency, $A_\perp^{(m)}$ is the perpendicular coupling of the nuclear spins and $\phi_{0m}$ the corresponding phase.
	The operators $\sigma_i$ act on the electron spin (NV center), $\sigma_i^{(m)}$ act on the mth nuclear spin.
	
	When we fulfil $\tau = \pi/\omega_L$ in the DD (e.g. XY-8) sequence, the
	sequence produces a modulation function
	{that modulates the NV $\sigma_z$ operator to $\sigma_z f(t)$ where}
	\begin{equation}
	f(t) = \frac{4}{\pi} \cos(\omega_L t) + \mathrm{rot.}
	\end{equation}
	and rot. denotes terms that vanish after the rotating-wave approximation in a frame rotating with the nuclear Larmor frequencies.
	Using
	\begin{equation}
	\cos^2(\omega_L t) = \frac{1}{2}\left(1+\cos(2\omega_L t)\right),
	\end{equation}
	
	we obtain the effective interaction Hamiltonian
	\begin{equation}\label{Heff}
	H_\mathrm{eff} = \sum \limits_m \frac{2A_\perp^{(m)}}{\pi} \sigma_z \otimes  \left(\sigma_x^{(m)} \cos({\phi_{0m}})  +  \sigma_y^{(m)} \sin({\phi_{0m}})  \right)
	\end{equation}

that will be the starting point for the following discussion.

}

We use ${k_0}={\mu_{n}} B_{s}/\hbar$ where $B_{s}$ is the
field generated by one of the $M$ auxiliary spins at the position of the NV center

\subsection{Transient super-Heisenberg scaling in measurement time}
In a first step we analyse the Fisher information scaling {with the} measurement time.
For this purpose we derive the probability $p_n$ of finding
the internal state of a spin-$1/2$ quantum sensor in the spin-down state in the n-th measurement in leading order
assuming the length $T_s$ of each instance of a magnetic field measurement is short and
if we apply these measurements every $\tau_m$.

{
	We simplify {to the case} that all couplings $A_\perp^{(m)} = k$ and all phases $\phi_{m}$ are equal, see the \BTF{appendix} for a more general discussion.
	Here $k_0={\mu_{n}} B_{s}/\hbar$ where $B_{s}$ is the field generated by one of the $M$ auxiliary spins at the position of the NV center. 
	In the described protocol the auxiliary spins gain a phase $\phi =    \delta \tau_m $ in each step,  where $\delta=2\pi(\nu-\nu_{loc})$ is the difference of precession frequency $\nu$ and local oscillator frequency $\nu_{loc}$.	
	For a nuclear spin with an already accumulated phase from $n$ cycles {equalling} $n \phi$
	the readout probability is calculated for the NV measurement in basis $Y${, i.e. the eigenbasis of the Pauli $,\sigma_y$ operator,} and NV preparation in $\ket{+}$. 
\begin{align}
p_n&-\frac{1}{2} = \mathrm{Tr} \left[\hat O_\mathrm{measurement} U \rho_n U^\dagger  \right]\nonumber
\\&= \mathrm{Tr} \left[\left(\frac{\sigma_y}{2}  \otimes \mathbbm{1}^{\otimes M} \right) U\nonumber
\right.\\&\left.
\left(\frac{\mathbbm{1}+\sigma_{x}}{2}  \otimes
\left(\frac{\mathbbm{1} + \cos(\phi n) \sigma_{x} +  \sin(\phi n) \sigma_y}{2}\right) \right)^{\otimes M} U^\dagger \right] \nonumber
\\&= \frac{i}{4} \left[ 
\left( \cos \frac{4 k_0 T_s}{\pi} - i \sin \frac{4 k_0 T_s}{\pi} \cos(\phi n) \right)^{M} \nonumber
\right.\\&\left.
- \left( \cos \frac{4 k_0 T_s}{\pi} + i \sin \frac{4 k_0 T_s}{\pi}  \cos(\phi n) \right)^{M}   \right].\label{noapprox}
\end{align}

For $M k_0 T_s \ll 1$ we can approximate
\begin{align}
&\left( \cos \frac{4 k_0 T_s}{\pi} \pm i \sin \frac{4 k_0 T_s}{\pi} \cos(\phi n) \right)^M \nonumber
\\
&\cong \exp\left(  \pm i M \sin \frac{4k_0 T_s}{\pi} \cos(\phi n)  \right)
\end{align}
to derive the signal

\begin{equation}
p_n = \cos^2\left(\frac{2M {k_0} T_s}{\pi}\cos\left(\phi n\right)-\frac{\pi}{4}\right).
\label{signal}
\end{equation}

Imperfect polarisation $P$ of the auxiliary spins can be incorporated via $k_0=\mu_{n} P
	B_{s}/\hbar$.
}

By eq. (\ref{signal}) we estimate the frequency $2\pi\nu = {\mu_{n}}B/\hbar$ and hence the
magnitude of the magnetic field $B$. For $N$ measurements, the achievable uncertainty in the
estimate of $\nu$ is obtained via the classical Fisher information
\begin{equation}
    {I_N = \sum_{n=1}^N \frac{1}{p_n(1-p_n)}\left(\frac{\partial p_n}{\partial \phi}\right)^2 \left(\frac{\partial \phi}{\partial \delta}\right)^2.}
\end{equation}
{We can describe decoherence processes in terms of a decay rate	$\gamma = \gamma_2+\gamma_b$
(see \BTF{appendix}), where $\gamma_2$ refers to $T_2$ processes on the nuclear spins and
$\gamma_b$ refers to measurement backaction from the quantum sensor. Then} the effective
coupling after $n$ measurements is $k_n = k_0 e^{-\gamma n}$ and we find
\begin{eqnarray}
	I_N(\gamma) =  \label{exact}  \sum_{n=1}^N \left(\frac{4 \tau_m M k_n T_s n}{\pi}\right)^2
	\sin^2 \phi n.
\end{eqnarray}
Under the assumptions $max[\gamma,\frac{1}{N}]\ll 2\pi\phi$ and $M k_0 T_s\ll 1$, i.e. when we
sample at least one full oscillation of the signal of frequency $\delta$, eq. (\ref{exact}) is well
approximated by
\begin{eqnarray}
    I_N(\gamma) &\cong& \frac{16 M^2 \tau_m^2 k_0^2 T_s^2}{\pi^2}\sum_{n=1}^N \frac{n^2}{2} e^{-2\gamma
    n}\left(1-\cos 2\phi n\right)\nonumber\\
    &\cong& \frac{2M^2 \tau_m^2 k_0^2 T_s^2}{\pi^2}\frac{1-e^{-2\gamma N}(1+2\gamma N(1+\gamma
    N))}{\gamma^3}.
    \label{approx}
\end{eqnarray}
For {$\gamma N \ll1$ (for $\gamma N < 0.6$ errors are smaller than 3\%)}
\begin{eqnarray}
    I_N(\gamma)
    \cong \frac{2 M^2 \tau_m^2 k_0^2 T_s^2}{\pi^2}(\frac{4N^3}{3}-2\gamma N^4 + \frac{6\gamma^2 N^5}{5})
\end{eqnarray}
and hence
\begin{eqnarray}
    \Delta B &\le& \sqrt{\frac{3\pi^2\hbar^2}{8{\mu_{n}}^2 M^2 \tau_m^2 k_0^2 T_s^2} \frac{1}{N^3}}.
\end{eqnarray}
As a result, for small $\gamma N$ our limited control procedure exhibits a scaling in the number
of measurements $N$ or, equivalently, the total measurement time $T = N \tau_m$ that exceeds the
standard HS of eq. (\ref{Heisenberg}) while the scaling in the number of particles $M$ achieves
the HL. {Note that unlike the case of interaction-based quantum metrology \cite{Boixo2007} this
super-Heisenberg scaling is not due to interactions between the auxiliary spins.} {Intuitively
the scaling can be understood to emerge due to the fact that, without noise, the last measurement
of the measurement record alone would already give quadratic Fisher information scaling, and the
linear number of intermediate measurements leads to a cubic scaling in total.}

\subsection{Asymptotic SQL scaling in measurement time}
However, {the super-Heisenberg scaling identified in the previous subsection} has to be transient and cannot persist for arbitrarily long times
as this would {be in} violation of the fundamental limit of sensitivity that is
imposed by the full control scheme in the absence of any noise.

For $\gamma_2=0$, the remaining contribution to the decay rate $\gamma$ is due to the
measurement backaction of the quantum sensor on the auxiliary spins which is negligible only for
$\gamma N\ll 1$. {Our measurement scheme then yields (see \BTF{appendix} for a derivation)}
\begin{equation}
    \gamma_b = \frac{4k_0^2T_s^2}{\pi^2}.
    \label{disturbance}
\end{equation}
Due to the measurement backaction, the signal weakens with increasing number of measurements $N$ and the rate of increase of the Fisher information slows. When determining the scaling in this
regime, a note of caution is in order as the calculation of the measurement backaction in
eq. (\ref{approx}) determines the Fisher information of the averaged density matrix of the auxiliary
spins. However, as we have access to and use all the intermediate measurements, the {correct Fisher information
of the protocol is obtained by weighted averaging over measurement trajectories}. For $\gamma N\gg 1$
this results in a scaling linear in $N$ (see also \cite{Pfender, Beige, cohen2019achieving}), as indicated analytically in
the \BTF{appendix} and numerically in figure \ref{Fig1}. For $M=100$ nuclei in an
initial product state (red data) the transient super-Heisenberg $I_N \propto N^3 \propto T^{3}$ scaling	
evolves into the shot noise scaling {(SQL)} of $I_N \propto N \propto T$ (blue asymptote) while eq. (\ref{approx}) {was calculated with the average density matrix and therefore}
would yield a constant. Remarkably, this linear scaling is independent of the initial state
and we can achieve the same scaling using a completely mixed initial state (orange triangles) \cite{Footnote3}. In the limit of small
interaction strength $k_0 T_s$ and decay, the asymptotic {($\gamma N \gg 1$)} value of the Fisher information can be estimated
to be
\begin{equation}\label{asymptote}
	I_N = \frac{\sin^4(4 k_0 T_s/\pi)}{16\left(\gamma_b + \gamma_2\right)^3} \frac{M^2}{2} \tau_m^2 N
\end{equation}

\begin{figure}
\centering
\includegraphics[width=.4\textwidth]{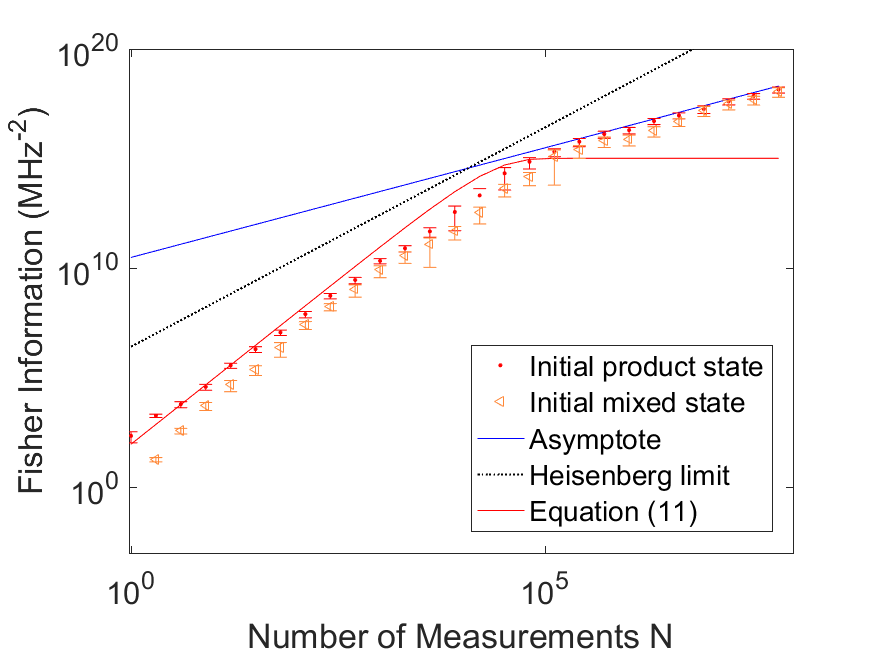}
\includegraphics[width=.4\textwidth]{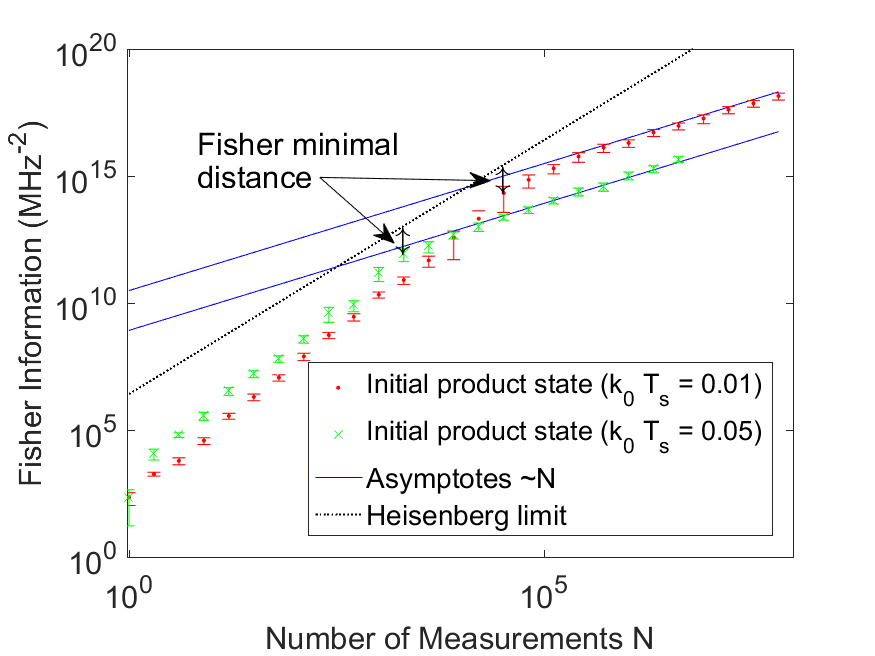}

\caption{Upper graph: Numerical Fisher information scaling (red: initial product state/ orange: initial mixed state), the analytical approximation for small $N$ eq. (\ref{approx}) (red line{, invalid for large N due to the average density matrix calculation being invalid for $\gamma N >1$}) and the
asymptotic {($\gamma N \gg 1$)} behaviour 
(blue) for {$\phi=0.7$ and} M=100 nuclear spins, each coupled with $k_0 T_s = 0.01$ to the NV center. The HL (black) for
$M+1$ spins is shown for comparison. The results are averaged over 96 (32 for mixed state) runs with $N=2^{26}$ measurements each.
{Lower graph: Same data (red) and results for
 $k_0 T_s = 0.05$ in green. The Fisher information is initially larger but dominated by backaction earlier, resulting in a smaller prefactor for the asymptotic regime.
{In both graphs and all other simulations the achievable Fisher information approaches the HL to within a factor $\lesssim 20$. This Fisher minimal distance is independent of the interaction strength $k_0 T_s$ and the number of nuclei $M$}, see \BTF{appendix}.}
}
\label{Fig1}
\end{figure}

\begin{figure}
	\centering
	\includegraphics[width=.4\textwidth]{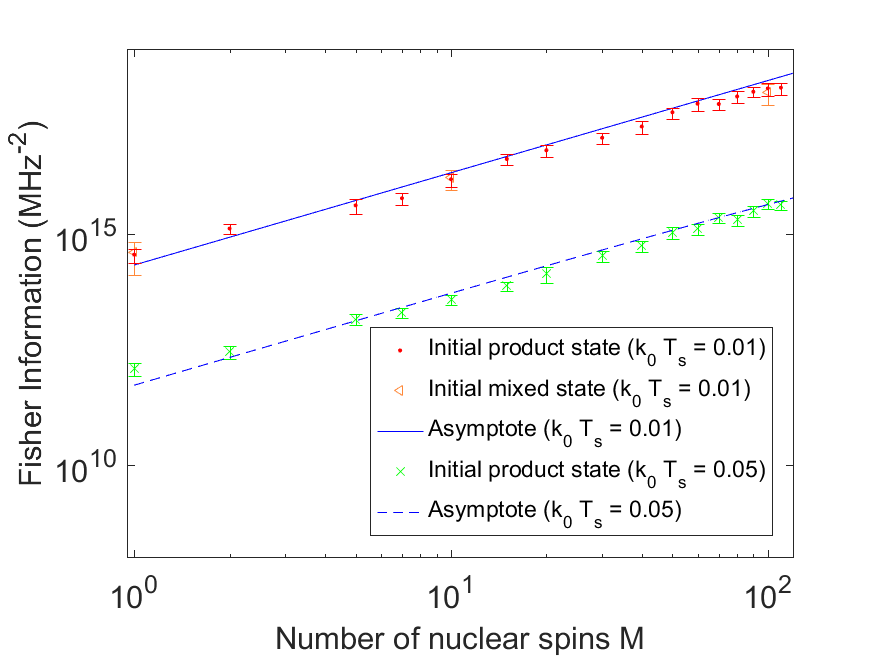}
	
\caption{
For the same parameters as in figure \ref{Fig1}: Numerical Fisher information scaling (red: initial product state/ orange: initial mixed state ($k_0 T_s = 0.01$ at $N = 2^{26}$ each)/ green: initial product state ($k_0 T_s = 0.05$ at $N = 2^{22}$))  with the number of nuclei $M$. The blue lines show the predicted scaling $\propto M^2$ from eq. (\ref{asymptote}) and Figure \ref{Fig1}.	
	}
	\label{Fig2}
\end{figure}

\subsection{Heisenberg scaling in particle number and relation to the Heisenberg limit}
{While the quadratic scaling with the number of nuclei $M$ is obvious for $\gamma N \ll 1$,
it also persists for $\gamma N \gg 1$ as confirmed in equation (\ref{asymptote}) and} in Fig.
\ref{Fig2} (see {Fig. \ref{Fig_Dec}} and the related discussion for details).
{In the full control scenario such a quadratic scaling can be traced back to the preparation
of an macroscopically entangled resource state of the form $(|0...0\rangle + |1...1\rangle)/\sqrt{2}$
including the quantum sensor and the auxiliary spins which contain one ebit of entanglement
\cite{Vedral98,Plenio05}. This entanglement is destroyed with the final measurement and represents 
the resource that is required to achieve HS.} {This is in sharp contrast with our scenario, 
which starts from an initial product state.} {Here the quadratic scaling {arises because 
the auxiliary spins are interrogated by a quantum sensor which results in a} readout operator that 
is not particle local, i.e. cannot be represented by a product of single particle operators {as 
would be the case when using a classical readout device}.}
{Indeed, every measurement applies a CPTP (completely positive and
trace-preserving) map to the auxiliary spins that is represented by the Kraus operators (see
also \cite{Pfender} for the case $M=1$ and our \BTF{appendix})}

\begin{equation}
U_\pm = \bra{\pm_y}  U  \ket{+}  = \frac{e^{-i \frac{2 k_0 T_s}{\pi} \sum \sigma_x^{(m)}} \mp i e^{i \frac{2 k_0 T_s}{\pi} \sum \sigma_x^{(m)}}}{2}
\end{equation}

{Crucially, this operator is diagonal in a basis different than the eigenbasis of the free
evolution operator - the latter being a tensor product of unitary evolutions on every auxiliary
spin. If the same held true for the readout operator, it is trivial to see that every spin would
be measured independently and therefor SQL scaling would apply. The readout operator here, however,
combines all particle states in a nontrivial, nonlocal, manner. This is similar to the metrological
advantage obtained for indistinguishable particles \cite{Braun} (section III).}
{In our work, this advantage is achieved by the sensor spin that, unlike a classical sensor,
allows to apply the same, particle non-local, CPTP map in every measurement.}
{We would like to stress that the entanglement build-up due to the backaction via a non-local
measurement operator does not contribute to the quadratic scaling with the number of auxiliary spins,
nor does it support the transient $N^3$ super-Heisenberg scaling.}

In the limit of large nuclear coherence
times $0 \approx \gamma_2 \ll \gamma_b$ the Fisher information approaches the HL
achievable under full control for $N_{opt}\approx \left({2 k_0 T_s}/{\pi}\right)^{-2}=1/
\gamma_b$ measurements after which the $N^3$-scaling {turns} into a scaling $\propto N$.
{{Note that the fundamental Heisenberg limit} is not violated as with decreasing interaction
	$k_0 T_s$ {both the backaction and the information gain per measurement decrease at the same rate thus compensating each other.}}
Remarkably, at this point the ratio of ultimate sensitivity under global control and the
limited control scheme used here only depends on the interaction strength and therefore can be tuned
to approach the HL {for $N_{opt}$ measurements to within {a Fisher minimal distance of} a factor independent of $T$, see figure \ref{Fig1}.
Furthermore it is natural to assume that this holds independently of $M$ as the Fisher information
of both the HL and the asymptote for $\gamma N\gg 1$ exhibit the same quadratic in $M$
scaling. This was confirmed numerically (see \BTF{appendix} for confirmation).

\section{Discussion} Metrology {assisted} by environmental spins have been considered
before, see e.g. \cite{GoldsteinCM+11,CappellaroGH+12}. There, however, the emphasis was placed
on spins that are strongly interacting with the quantum sensor and the measurement protocol creates
a joint entangled state of the quantum sensor and the auxiliary spins which then evolves
for some time followed by an inversion of the entangling operation and the subsequent measurement of
the state of the quantum sensor. In this approach the HS is achieved in the number of
{\em strongly} coupled auxiliary spins while we assume no such spins in our set-up. Furthermore, this protocol suffers from the drawback that it is fundamentally limited by the coherence
time of the quantum sensor and hence does not take full advantage of the long coherence time of the
auxiliary spins. In contrast, the measurements in our protocol can be made shorter than the
coherence time without adversely affecting the achievable sensitivity.

Besides the theoretical interest in the novel scaling regimes, we stress that the proposed
scheme employing auxiliary spins under limited control provides enhanced sensitivity as compared to the quantum sensor alone. This advantage
is the result of two processes. First, the transduction of the static magnetic field to a time-dependent
Larmor precession which is then detected by the quantum sensor facilitates the use of dynamical decoupling
schemes to filter out noise without adversely affecting the signal. Secondly, as each auxiliary spin contributes
to the signal, the overall signal strength scales with the number of spins and hence leads to a considerable
signal enhancement. 
{Remarkably, magnetometry schemes such as atoms in gas cells which are probed independently 
by a classical field lead to a $M^{-1}$ scaling of the variance with the particle number $M$. In 
sharp contrast, it is the transduction of the signal to a quantum sensors, e.g. an NV center, which 
results in a particle non-local measurement which causes the $M^{-2}$ Heisenberg scaling. This suggests 
a practical route for enhancing the measurement capacity of gas cell magnetometers.}
Furthermore, for an NV center as quantum sensor, even when considering nuclei with their
small magnetic moment as auxiliary spins, we may obtain an increased sensitivity. To this end, let us consider
the $|m=0\rangle \leftrightarrow|m=+1\rangle$ transition of an NV center in an external magnetic field $B$
and assume that the NV center is dominated by pure dephasing which results in a coherence time {$T_2^{(NV)}$}.
	For perfect readout efficiency, the optimal interrogation scheme yields 
	\begin{equation}
{\Delta B = \sqrt{\frac{2e\hbar^2}{\mu_e^2T_2^{(NV)}T}}
	= \sqrt{\frac{4e\hbar^2}{\mu_e^2 N{T_2^{(NV)}}^2}}} 
	\end{equation}	
\cite{HuelgaMP+97} where {$N=2T/T_2^{(NV)}$}. 
{The maximum of equation (\ref{asymptote}) that takes the form $\gamma_b^2/(\gamma_b+\gamma_2)^3$ is obtained by choosing
	$\gamma_b = 2\gamma_2$.
This allows to compare}
with our indirect measurement scheme using $M$ hydrogen nuclear spins and assuming
$\left(\frac{\mu_e}{{\mu_{n}}}\right)^2 \frac{T_2^{(NV)}}{T_2^{n}} \approx 10^3$ we find that for
\begin{equation}
	{M > \sqrt{\frac{27}{4e}\frac{\mu_e^2  T_2^{(NV)}  }{\mu_n^2  T_2^{(n) }}} \approx 50}
\end{equation}
the auxiliary spin assisted sensor outperforms the bare NV center.
{Nuclear spins couple more weakly to both noise and signal due to smaller gyromagetic ratio resulting in typically larger coherence times $\frac{\mu_e}{{\mu_{n}}} \approx \frac{T_2^{n}}{T_2^{(NV)}} $, so similar results apply for other systems.}
While our protocol makes use of a far
smaller magnetic moment compared to even a single electron spin $M \mu_n \ll \mu_e$, this is compensated
by the longer coherence time and the possibility to measure during the signal accumulation.

{Furthermore the obtained expression also highlights the advantage of the $M^{-2}$ Heisenberg scaling over the SQL that individual measurements on the nuclear spins give, providing higher Fisher Information for $M>27/4e\approx 2.5$.}

Finally, we note that our analysis also covers the case $M=1$ corresponding to
the detection of the Larmor frequency of a single nuclear spin via an NV center.  Super-Heisenberg scaling
applies for as long as the measurement backaction is weak. This applies for distant nuclear spins or
for measurements that are designed to be weak, i.e. not obtaining a full bit of information in each
single measurement.

{\em Conclusions --} We have examined metrology in a realistic setting of limited control and found transient
super-Heisenberg scaling in the total measurement time and a metrological precision approaching that
of the same number of particles under full experimental control. This is despite the absence of initial
entanglement in the system. In fact, in this scheme entanglement emerges only with increasing number
of measurements and adversely affects the metrological scaling. Furthermore, the proposed set-up, which employs
auxiliary spin under limited control, also represents an hybrid sensor that may outperform a bare quantum
sensor thus providing new design principles for quantum sensors.

{\em Acknowledgements --} The authors thank Liam McGuinness and Jan F. Haase for discussions 
and comments on the manuscript. This work was supported by the ERC
Synergy Grant BioQ, the EU projects AsteriQs, HYPERDIAMOND, the BMBF projects NanoSpin and DiaPol,
the DFG CRC 1279 and the DFG project 414061038. The authors acknowledge support by the state of
Baden-W{\"u}rttemberg through bwHPC and the German Research Foundation (DFG) through grant no
INST 40/467-1 FUGG (JUSTUS cluster). The first results of this work have been presented at the
Workshop On Quantum Metrology, 22nd - 23rd June 2017 in Ulm, Germany.


\begin{thebibliography}{38}
\bibitem[FN1()]{FN1}
	Our considerations are easily extended to the case of an arbitrary polarisation
	as the signal and hence the sensitivity is directly proportional to the
	polarisation level.
\bibitem[Foo()]{Footnote3}
	See also \cite{GoreckaPL+18} for the related observation that for metrology in
	non-markovian environments the zeno-scaling can be achieved with a mixed
	initial state.
\end{thebibliography}

\widetext

\newpage
\appendix

{\Large Appendix: Limited-control metrology approaching the Heisenberg limit without entanglement preparation}

\section{Derivation of the signal for few measurements}

{This section presents a more general calculation of equations (6-8) of the main text.}
For a nuclear spin (initial state described by polarisation P) with an already accumulated phase from $n$ cycles $\phi_{1m} =    \delta \tau_m n + \phi_0 $ the readout probability is
(NV measurement in basis $X c_\alpha + Y s_\alpha$ and NV preparation in $X$)
\begin{align}
p_n-\frac{1}{2} &= \mathrm{Tr} \left[\hat O_\mathrm{measurement} U \rho_n U^\dagger  \right]
\\&= \mathrm{Tr} \left[\left(\frac{\sigma_x \cos\alpha + \sigma_y \sin\alpha}{2}  \otimes \mathbbm{1}^{\otimes M} \right) U \left(\frac{\mathbbm{1}+\sigma_{x}}{2}  \otimes
\prod\limits_{m=1}^M \left(\frac{\mathbbm{1} +P \cos\phi_{1m} \sigma_{x}^{(m)} + P \sin\phi_{1m} \sigma_y^{(m)}}{2}\right) \right) U^\dagger \right]
\\&= \frac{1}{4} \left[ \cos\alpha \left(\prod\limits_{m=1}^M\left( \cos \frac{4 A_\perp^{(m)} T_s}{\pi} - i \sin \frac{4 A_\perp^{(m)} T_s}{\pi} P \cos\phi_{m} \right) + \prod\limits_{m=1}^M\left( \cos \frac{4 A_\perp^{(m)} T_s}{\pi} + i \sin \frac{4 A_\perp^{(m)} T_s}{\pi} P \cos\phi_{m} \right)\right)
\right.\\&\left.
+i \sin \alpha \left(\prod\limits_{m=1}^M\left( \cos \frac{4 A_\perp^{(m)} T_s}{\pi} - i \sin \frac{4 A_\perp^{(m)} T_s}{\pi} P \cos\phi_{m} \right) - \prod\limits_{m=1}^M\left( \cos \frac{4 A_\perp^{(m)} T_s}{\pi} + i \sin \frac{4 A_\perp^{(m)} T_s}{\pi} P \cos\phi_{m} \right)\right) \right]\label{noapprox}
\end{align}
where $\phi_m = \phi_{1m}-\phi_{0m}$.

For $\sum\limits_{m=1}^M  4 A_\perp^{(m)} T_s/\pi \ll 1$ we can approximate

\begin{align}
\prod\limits_{m=1}^M\left( \cos \frac{4 A_\perp^{(m)} T_s}{\pi} \pm i \sin \frac{4 A_\perp^{(m)} T_s}{\pi} P \cos\phi_{m} \right)
&=\prod\limits_{m=1}^M \exp\left(  \pm i \sin \frac{4 A_\perp^{(m)} T_s}{\pi} P \cos\phi_{m}   +O\left(\frac{4 A_\perp^{(m)} T_s}{\pi}\right)^2  \right)
\\
&\cong \exp\left(  \pm i \sum\limits_{m=1}^M \sin \frac{4 A_\perp^{(m)} T_s}{\pi} P \cos\phi_{m}  \right)
\end{align}

to derive the signal

\begin{align}
p_n
&= \frac{1}{2} + \frac{1}{2}\cos\left(  \sum\limits_{m=1}^M \sin \frac{4 A_\perp^{(m)} T_s}{\pi} P \cos\phi_{m} - \alpha \right)  = \cos^2 \left(  \sum\limits_{m=1}^M \sin \frac{2 A_\perp^{(m)} T_s}{\pi} P \cos\phi_{m} - \frac{\alpha}{2} \right).
\end{align}


{The generalisation of this case described in the main text to different coupling $A_\perp^{(m)} \ne A_\perp^{(m)}$ can be described by an effective coupling.
	Different phases $\cos\phi_{m_1} \ne \cos\phi_{m_2}$ can produce additional features, but in the asymptotic case these effects are irrelevant as the initial state becomes less important, see next section.}

\section{Derivation of the Fisher Information from the probabilities for a full measurement record}

The outcome of the kth measurement is denoted by
every individual $X_k \in \{0=+,1=-\}$, $X^k$ is a measurement record of the form $\{1,0,1,1,0,1,...\}$ with k components, $X_l$ is the lth component of it and $\beta \equiv 2 A_x T_s/\pi$ is the coupling achieved by the XY-sequence.
Furthermore $U_\phi = \exp\left(-i \phi \sum \sigma_z^{(m)}/2\right)$ and 
\begin{equation}
U_\pm = \bra{\pm_y}  U  \ket{+}  = \frac{e^{-i \beta \sum \sigma_x^{(m)}} \mp i e^{i \beta \sum \sigma_x^{(m)}}}{2}
\end{equation}
with the coupling from the first section $U = \exp(-i \beta \sigma_z^{NV} \sum \sigma_x^{(m)})$.
Here we assumed all coupling constants to be equal, however different $\beta_m$ don't change the structure of the result.


Each measurement probability can be described by
\begin{equation}
p_\pm = \mathrm{Tr} \left[ (\ket{\pm_y}\bra{\pm_y}\otimes \mathbbm{1}) U U_\phi (\ket{\pm}\bra{\pm}\otimes\rho_{N-1})U_\phi^\dagger U^\dagger\right] = \mathrm{Tr}\left[U_{X_1}U_\phi \rho_0 U_\phi^\dagger U_{X_1}^\dagger\right]
\end{equation}
and evolves the nuclear state to
\begin{equation}
\rho_{\pm} = \frac{1}{p_\pm}\mathrm{Tr_{NV}} \left[U_{X_1}U_\phi \rho_0 U_\phi^\dagger U_{X_1}^\dagger\right].
\end{equation}

The probability for a measurement record $X^k$ can be described as

\begin{align}
{p_{X^k}} &= p_{X_1} p_{X_2|X_1} ... p_{X_{N-1}|X_{N-2}...X_1}  p_{X_N|X_{N-1}}
\\&= \mathrm{Tr}\left[U_{X_1}U_\phi \rho_0 U_\phi^\dagger U_{X_1}^\dagger\right] \frac{\mathrm{Tr}\left[U_{X_2}U_\phi U_{X_1}U_\phi \rho_0 U_\phi^\dagger U_{X_1}^\dagger U_\phi^\dagger U_{X_2}^\dagger\right]}{\mathrm{Tr}\left[U_{X_1}U_\phi \rho_0 U_\phi^\dagger U_{X_1}^\dagger\right]}...
\\&= \mathrm{Tr}\left[ \prod\limits_{k=1}^{N} \left(U_{X_k}U_\phi\right) \rho_0 \left(\prod\limits_{k=1}^{N} \left(U_{X_k}U_\phi\right)\right)^\dagger   \right].
\end{align}

Each part of the sum contributes roughly $2^{-N}$, sum over all contributions is 1. To analyse the effect of these operators, we apply them on a permutation-invariant product state

\begin{equation}
U_\phi \left(\frac{a \mathbbm{1} + b \sigma_x + c \sigma_y + d \sigma_z}{2}\right)^{\otimes M}  U_\phi^\dagger
= \left(\frac{a \mathbbm{1} + \left(b c_\phi - c s_\phi \right) \sigma_x + \left(c c_\phi + b s_\phi \right) \sigma_y + d \sigma_z}{2}\right)^{\otimes M}
\end{equation}

\begin{align}
4  U_\pm \left(\frac{a \mathbbm{1} + b \sigma_x + c \sigma_y + d \sigma_z}{2}\right)^{\otimes M}  U_\pm^\dagger
&= \left(\frac{a \mathbbm{1} + b \sigma_x + \left(c c_{2\beta} + d s_{2\beta} \right) \sigma_y + \left(d c_{2\beta} -  c s_{2\beta} \right) \sigma_z}{2}\right)^{\otimes M}  \label{eq9}
\\&+
\left(\frac{a \mathbbm{1} + b \sigma_x + \left(c c_{2\beta} - d s_{2\beta} \right) \sigma_y + \left(d c_{2\beta} +  c s_{2\beta} \right) \sigma_z}{2}\right)^{\otimes M}\label{eq10}
\\&\pm i \left[  \left(\frac{\left(a c_{2\beta} +i b s_{2\beta} \right) \mathbbm{1} + \left(b c_{2\beta} +i a s_{2\beta} \right) \sigma_x + c \sigma_y + d \sigma_z}{2}\right)^{\otimes M}
\right.
\\&\left. \hspace{.5 cm}  -\left(\frac{\left(a c_{2\beta} -i b s_{2\beta} \right) \mathbbm{1} + \left(b c_{2\beta} -i a s_{2\beta} \right) \sigma_x + c \sigma_y + d \sigma_z}{2}\right)^{\otimes M}
\right].
\end{align}

It is very difficult to calculate the full expression because every measurement multiplies the number of terms by 4.
So we want to find the relevant terms for the Fisher Information for different limits.
For a single nucleus M=1, only two terms are created every measurement and $d \sigma_z$ can be neglected. Therefore we simplify to

\begin{align}
4  U_\pm \left(\frac{a \mathbbm{1} + b \sigma_x + c \sigma_y}{2}\right)  U_\pm^\dagger
&= 2\left(\frac{a \mathbbm{1} + b \sigma_x + c c_{2\beta} \sigma_y}{2}\right)
\pm   2\left(\frac{- b s_{2\beta}\mathbbm{1} - a s_{2\beta} \sigma_x + c \sigma_y}{2}\right)
\\&= 2\mathcal{A} [\rho] + 2\mathcal{B} [\rho].
\end{align}

The $\cos(2\beta)$ for the $c$ coefficient produces the backaction-induced decay $\gamma_b$. We can approximate 
\begin{equation}
\mathcal{A} [\rho] = \left(\frac{a \mathbbm{1} + b \sigma_x + c c_{2\beta} \sigma_y}{2}\right)   \approx
\left(\frac{a \mathbbm{1} + (1+c_{2\beta})/2 \left(b \sigma_x + c \sigma_y\right)}{2}\right).
\end{equation}
which reduces the bloch vector according to
\begin{equation}
\left(\frac{1+c_{2\beta}}{2}\right)^k = \exp\left(k \log \left(\frac{1+c_{2\beta}}{2}\right)\right)
\approx \exp\left(k \log \left(1 + \beta^2 \right)\right)
\approx \exp\left(-k \beta^2 \right).
\end{equation}

This is valid because the higher orders will be negligible in the further calculation.
T2 processes have a similar effect, why we define the decay of population in the x-y-plane with
\begin{equation}
\gamma = -\log\frac{1+c_{2\beta}}{2} + \frac{\tau_m}{T_2^{(nuc)}}  \approx \beta^2 + \frac{\tau_m}{T_2^{(nuc)}}
\end{equation}
where $\tau_m$ is the time for each of the N repetitions.

When starting with $\rho_0 = (\mathbbm{1}+\sigma_x)/2$ can expand the probability for a  N measurement record $X^N$ as

\begin{align}
2^N p_{X^N} &\approx 1 + s_{2\beta} \sum\limits_{l=1}^N (-1)^{X_l} \exp(- \gamma l) \cos(l \phi)
\\&+ s_{2\beta}^2 \sum\limits_{1 \leqslant l_1 < l_2 \leqslant N} (-1)^{X_{l_1}}(-1)^{X_{l_2}} \exp(- \gamma (l_2-l_1)) \cos((l_2-l_1) \phi)
\\&+ s_{2\beta}^3 \sum\limits_{1 \leqslant l_1 < l_2 < l_3 \leqslant N} (-1)^{X_{l_1}}(-1)^{X_{l_2}}(-1)^{X_{l_3}} \exp(- \gamma (l_3-l_2+l_1)) \cos((l_3-l_2+l_1) \phi)
\\&+...
\end{align}

In a first step we
calculate the Fisher Information
\begin{align}
I_N &= \sum\limits_{X^N} \frac{1}{p_{X^N}} \left(\frac{\partial p_{X^N}}{\partial \phi}\right)^2 \left(\frac{\partial \phi}{\partial \delta}\right)^2
\end{align}
in the limit $\gamma N \ll 1$
using a geometric series 

\begin{align}
\frac{1}{\tau_m^2}  I_N &= 2^{-N} \sum\limits_{X^N} \sum\limits_{k=0}^\infty \left(-s_{2\beta} \sum\limits_{l=1}^N (-1)^{X_l} \exp(- \gamma l) \cos(l \phi)- s_{2\beta}^2...   \right)^k
\left(-s_{2\beta} \sum\limits_{l=1}^N (-1)^{X_l} \exp(- \gamma l) l\sin(l \phi) - ...     \right)^2.
\end{align}

As we average over all $(-1)^{X_l} = \pm 1$, only terms with an even number of all $(-1)^{X_l}$ contribute.
The first order ($\gamma N \ll 1$, $k=0$) results in the $N^3$ scaling that is discussed in the main text.

\begin{align}
\frac{1}{\tau_m^2}  I_N &= s_{2\beta}^2 \sum\limits_{l=1}^N \exp(-2 \gamma l) l^2\sin^2(l \phi)
\\&\approx \frac{s_{2\beta}^2}{2} \int\limits_{0}^N \mathrm{d}l \exp(-2 \gamma l) l^2 (1-\cos(2 l \phi))
\\&\approx \frac{s_{2\beta}^2}{2} \int\limits_{0}^N \mathrm{d}l \exp(-2 \gamma l) l^2
\\&= \frac{s_{2\beta}^2}{2}\frac{e^{-2 \gamma N}(-2\gamma N(\gamma N+1)-1)+1}{4\gamma^3} \approx \frac{s_{2\beta}^2}{2} \left[\frac{N^3}{3} - \frac{\gamma N^4}{2}\right].
\end{align}

For $\gamma N > 1$ the
geometric series is not valid anymore and many higher orders in $l$ need to be considered. To show that terms linear in $N$ exist, we consider the approximated second order ($1/p_X^N\approx 2^N$)

\begin{align}
\frac{1}{\tau_m^2}  I_N &= s_{2\beta}^4 \sum\limits_{1 \leqslant l_1 < l_2 \leqslant N} \exp(-2 \gamma (l_2-l_1)) \sin^2((l_2-l_1) \phi) (l_2-l_1)^2
\\&\overset{l=l_2-l_1}{=} s_{2\beta}^4 \sum\limits_{1 \leqslant l \leqslant N} (N-l) \exp(-2 \gamma l) \sin^2(l \phi) l^2
\\&\approx \frac{s_{2\beta}^4}{2} \int\limits_{0}^N \mathrm{d}l (N-l) \exp(-2 \gamma l) l^2 (1-\cos(2 l \phi))
\\&\approx \frac{s_{2\beta}^4}{2} \int\limits_{0}^N \mathrm{d}l (N-l) \exp(-2 \gamma l) l^2
\\&= \frac{s_{2\beta}^4}{2}\frac{e^{-2 \gamma N}(2\gamma N(\gamma N+2)+3) + 2\gamma N - 3}{8\gamma^4} \approx \frac{s_{2\beta}^4}{2} \frac{N}{4\gamma^3}.
\end{align}

The numerically obtained prefactor from the main text is smaller by a factor 2 in case of $M=1$ and a factor $\left(c_{2\beta}^{M-1}\right)^2 M^2/4$ for $M>1$.
While the first factor 2 is likely to originate from higher order contributions (the dominant order in $l$ depends on $N$), the difference for higher $M$ can be explained by the additional terms that arise in the calculation.
Many terms like in
\begin{align}
2^N p_{X^N} &\approx 1 + \frac{i}{2}\sum\limits_{l=1}^N (-1)^{X_l} \left[\left(c_{2\beta} - i s_{2\beta}  \exp(- \gamma l) \cos(l \phi)\right)^M
- \left(c_{2\beta} + i s_{2\beta} \exp(- \gamma l) \cos(l \phi)\right)^M\right]
\\&+\frac{i^2}{4} \sum\limits_{1 \leqslant l_1 < l_2 \leqslant N} (-1)^{X_{l_1}}(-1)^{X_{l_2}}\left[\left(c_{2\beta}^2 - s_{2\beta}^2  \exp(- \gamma (l_2-l_1)) \cos((l_2-l_1) \phi) + i\alpha \right)^M +... \right]
\\&+...
\end{align}
will produce roughly the same Fisher information, in particular the first order scales as expected. The derivative gives a factor $M^2$, and the leading order has a factor $\left(c_{2\beta}^{M-1}\right)^2$. The additional factor 1/2 will arise from averaging random phases $\alpha$.

{The scaling $\propto \gamma^{-3}$ was tested numerically in figure \ref{Fig_Dec}. The simulation results seem to deviate in a regime $\gamma_2\approx \gamma_b$ by a factor of 2. This can be explained by higher order terms being affected more by $\gamma_2$. As a result, lower order terms with the expected scaling dominate.}

{Remarkably these asymptotic results are independent of the initial state of the nuclei. Therefore different phases $\cos\phi_{m_1} \ne \cos\phi_{m_2}$ can be transformed to a basis with equal phases and a different initial state, which yields the same result as for $\cos\phi_{m_1} = \cos\phi_{m_2}$ $\forall m_1,m_2$.}

\begin{figure}
	\centering
	
	\includegraphics[width=.4\textwidth]{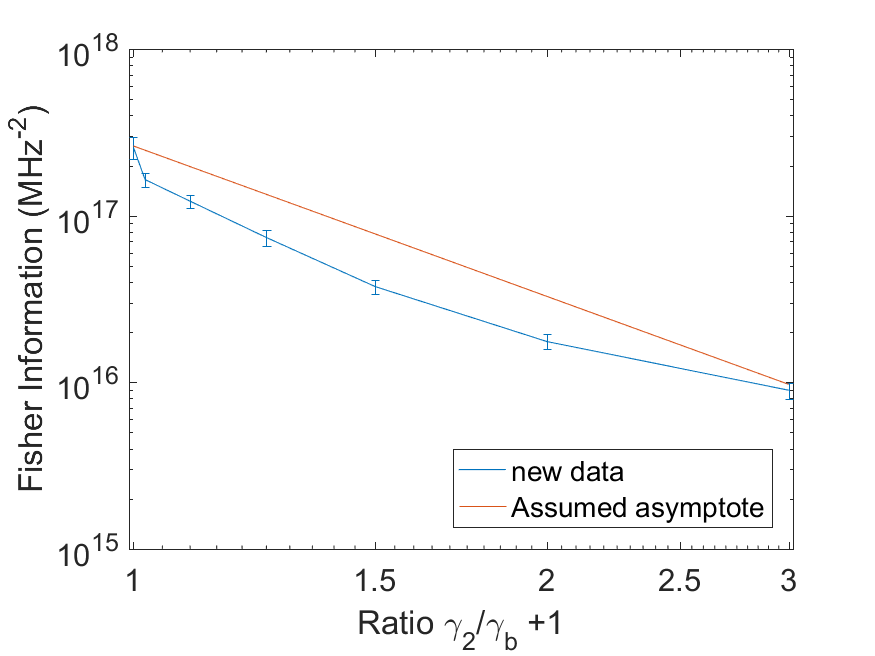}
	
	\caption{
		{For M=10 nuclei: Fisher information after $2^{24}$ measurements for $\beta = 0.01\times 2/\pi$ for different $\gamma_2$ averaged over 192 runs compared to a $\gamma^{-3}$ curve with the numerically obtained prefactor.
		}
	}
	\label{Fig_Dec}
\end{figure}

\section{Relation to the Heisenberg limit}

We numerically investigated the minimum difference between our protocol and the Heisenberg limit, which can only be achieved given full control over the nuclei in absence of decoherence $\gamma_2 = 0$.

{
	Figure \ref{Fig2SI} shows the maximum ratio between the Fisher Information of the investigated protocol and the Heisenberg limit. For different coupling strength the ratio is independent of the number of nuclei.
	Note that the peaks are due to the monte carlo simulation. This is confirmed by the curves on the right hand side of figure \ref{Fig2SI}, where the maxima are found at $N \approx \beta^{-2}$ as expected. }

\begin{figure}
	\centering
	
	\includegraphics[width=.4\textwidth]{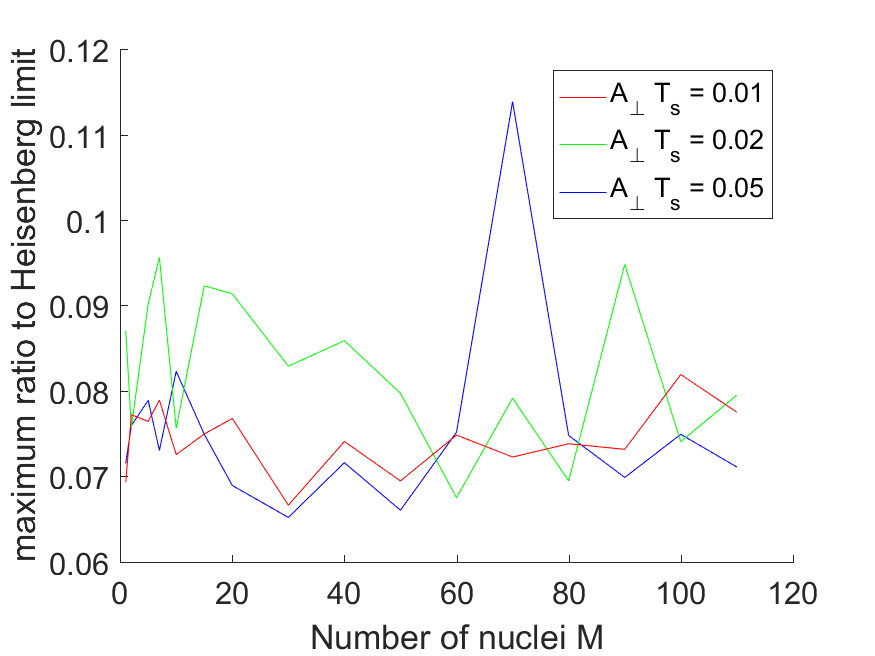}
	\includegraphics[width=.4\textwidth]{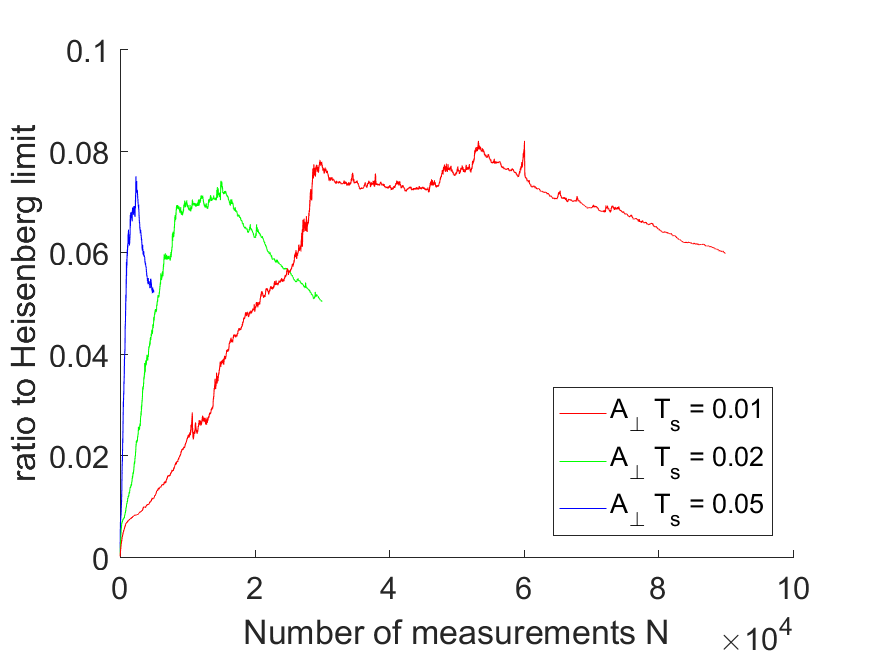}

	\caption{
		{Left: Maximum of the ratio between the Fisher information of our protocol and the Heisenberg limit 
			{for different coupling strength. The results do not depend on the number of nuclei $M$, but are dominated by variations originating from the 9600 averages in the monte carlo simulation. Right: Curves for $M=100$ nuclei: The highest value is reached at $N \approx \beta^{-2}$ as expected. Artefacts of the simulation are clearly visible as for an infinite number of repetitions smooth curves are expected.}
		}
	}
	\label{Fig2SI}
\end{figure}

\section{Simulation}\label{invsim}

The normal simulation (without making use of the permutation invariance) repeats the following steps (after initializing the nuclear spins to $\ket{\psi_0}=\ket{+}^{\otimes M}$, $\rho_0=\ket{\psi_0}\bra{\psi_0}$)

\begin{enumerate}
	\item Simulate nuclear spin evolution with the operator
	\begin{equation}
	U_\mathrm{free} = \exp \left(-i  \delta \tau_m \sum_m \sigma_{z}^{(m)}/2\right)
	\end{equation}
	where $\tau_m$ is the {time between two measurements.}
	\item Determine probability to measure the NV in $\ket{+_y}$ after preparing it in $\ket{+}$ and evolving it with the nuclear spins according to (\ref{Heff}) by
	\begin{align}
	p &= \mathrm{Tr}\left[ \ket{+_y}\bra{+_y}\otimes \mathbbm{1}^{\otimes M} U \ket{+}\bra{+}\otimes \rho_n U^\dagger  \right]
	= \mathrm{Tr}\left[ U_+ \rho_n U_+^\dagger \right]
	\\&= \bra{\psi_n} U_+^\dagger U_+ \ket{\psi_n} \text{(for pure states)}\label{eq6}
	\end{align}
	where $U_+ = \bra{+_y}U\otimes \mathbbm{1}^{\otimes M}  \ket{+}$
	\item Probabilistically choose result according to $p$, save result and evolve accordingly including normalisation
	$\rho_{n+1}= \mathcal{N} U_{+/-} \rho_n U_{+/-}^\dagger$.
\end{enumerate}

By using the subspace resulting from the symmetry in the case of many spins with equal coupling strength, many spins can be simulated efficiently, as this subspace has dimension $M+1$ instead of $2^M$.

The Fisher Information
\begin{equation}
I_N = \sum\limits_{X^N} p_X\frac{1}{p_X^2} \left(\frac{\partial p_X}{\partial \delta}\right)^2
\end{equation}
was calculated numerically for many different runs evolving $\rho$ following the recipe above to determine $p_X$
Evolving $\rho_{2/3}$ according to the same measurement outcomes as $\rho$, but with a different evolution parameter $\delta \pm \mathrm{d}\delta$ allows to determine
\begin{equation}
\left(\frac{\partial p_X}{\partial \delta}\right) = \frac{p_X(\delta + \mathrm{d}\delta)- p_X(\delta - 
	\mathrm{d}\delta)}{2\mathrm{d}\delta}
\end{equation}
for many measurement records. After calculating the Fisher Information for every measurement record, the average 
and standard deviation can be obtained

The accuracy is limited by the Fisher Information due to the {Cramer-Rao bound}

\begin{equation}
\delta \omega_N \ge \frac{1}{\sqrt{I_N}}.   
\end{equation}

For pure states, the Logarithmic negativity can be simplified to an expression depending on the Schmidt coefficients $\alpha_i$:
\begin{align}
LN(\ket{\Psi}\bra{\Psi}) = 2 \log \left(\sum\limits_i \alpha_i\right),
\end{align}
which can be calculated for considerably larger systems than the partial trace.

{
	In order to obtain some insights into the entanglement buildup and its potential role as a
	resource in the metrology scheme, we use the logarithmic negativity \cite{Plenio05} as a
	quantifier of the entanglement between one of the auxiliary spins with the remaining $M-1$
	spins and between equal bi-partitions of the auxiliary spins. While the entanglement between
	the nuclei builds up to a steady state after a time $~1/\gamma_b$,
	it does not contribute to the quadratic
	scaling with the number of auxiliary spins as this effect is related to the readout process, nor does it support the $N^3$ super-Heisenberg scaling.
	However, destroying the entanglement after every measurement would lead to lower prefactor in the asymptotic SQL scaling as it inevitably leads to destruction of information.
	Figure \ref{Fig3SI} shows this buildup of entanglement of a scale of $N_{opt}\approx \left({2 k_0 T_s}/{\pi}\right)^{-2}=1/
	\gamma_b$.
}

\begin{figure}
	\centering
	\includegraphics[width=.4\textwidth]{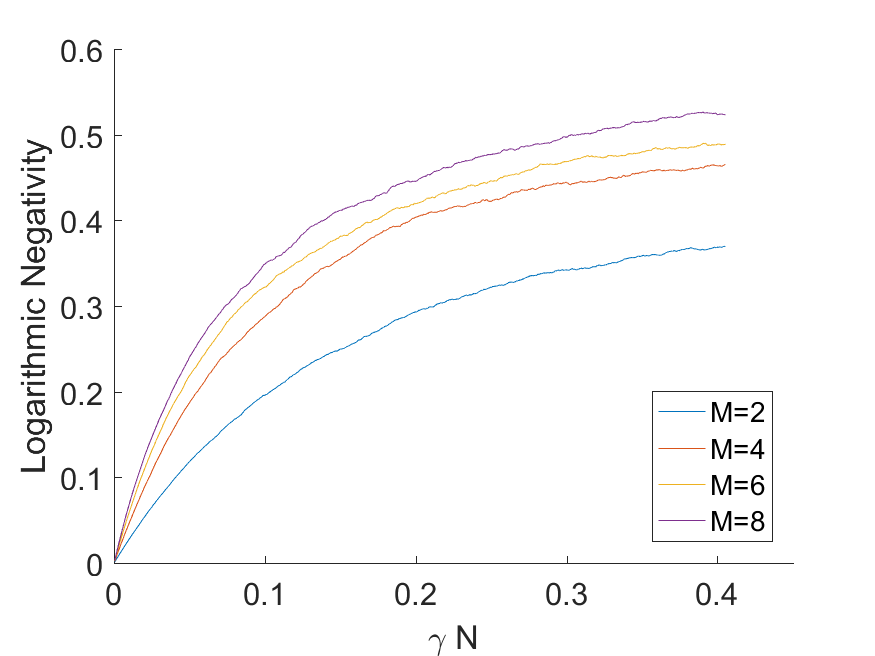}
	\includegraphics[width=.4\textwidth]{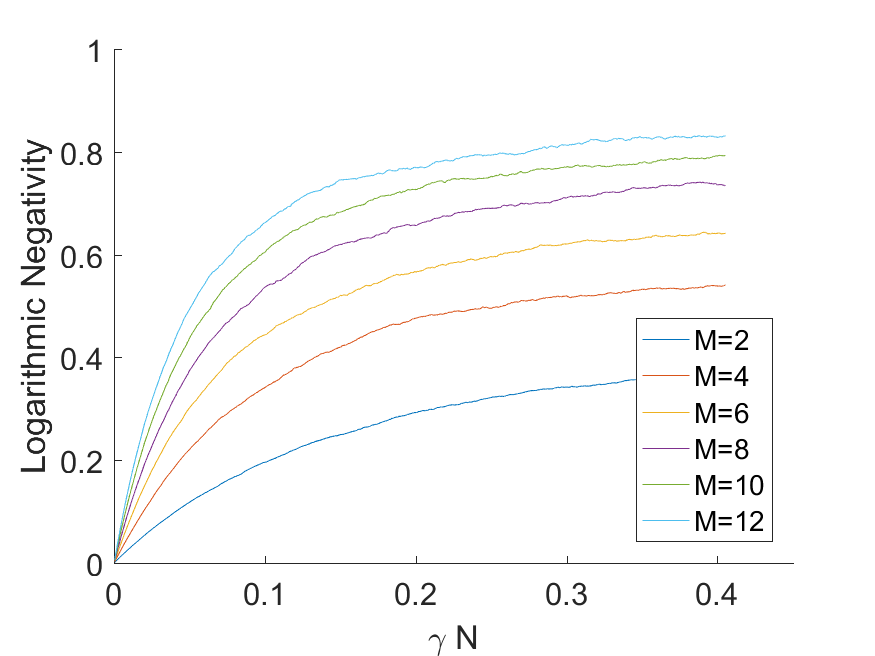}
	\caption{
		{Logarithmic negativity for $M$ spins, each spin coupled with $ \beta = 0.01 \times 2/\pi$ with negligible decay $\gamma_2 =0$. The left graph shows the entanglement of one of the
			auxiliary spins with the remaining M-1 spins and the right graph shows entanglement in an equal bi-partition
			of the auxiliary spins.
			The results are averaged over 2000 runs.}
	}
	\label{Fig3SI}
\end{figure}

\newpage

\end{document}